\pdfoutput=1
\documentclass[preprint,12pt,authoryear]{elsarticle}
\setcitestyle{numbers,square,sort&compress,comma}

\usepackage{amssymb}
\usepackage{amsmath}
\usepackage{subfigure}

\usepackage{booktabs}
\usepackage{diagbox} 
\usepackage{multirow}
\usepackage{geometry}

\begin{document}

\begin{frontmatter}

\title{TransCC: Transformer Network for Coronary Artery CCTA Segmentation}

\cortext[cor1]{Corresponding author}
\author[1,2]{Chenchu Xu}
\author[1]{Meng Li}
\author[3]{Dong Zhang\corref{cor1}}
\ead{zhangdong9612@gmail.com}
\author[1]{Xue Wu\corref{cor1}}
\ead{wuxue_2001@163.com}

\address[1]{Artificial Intelligence Institute, Anhui University, Hefei, China}

\address[2]{Institute of Artificial Intelligence, Hefei Comprehensive National Science Center, Hefei, China}

\address[3]{Department of Electrical and Computer Engineering, University of British Columbia, Vancouver, Canada}

\begin{abstract}
The accurate segmentation of Coronary Computed Tomography Angiography (CCTA) images holds substantial clinical value for the early detection and treatment of Coronary Heart Disease (CHD). The Transformer, utilizing a self-attention mechanism, has demonstrated commendable performance in the realm of medical image processing. However, challenges persist in coronary segmentation tasks due to (1) the damage to target local structures caused by fixed-size image patch embedding, and (2) the critical role of both global and local features in medical image segmentation tasks.To address these challenges, we propose a deep learning framework, TransCC, that effectively amalgamates the Transformer and convolutional neural networks for CCTA segmentation. 
Firstly, we introduce a Feature Interaction Extraction (FIE) module designed to capture the characteristics of image patches, thereby circumventing the loss of semantic information inherent in the original method. Secondly, we devise a Multilayer Enhanced Perceptron (MEP) to augment attention to local information within spatial dimensions, serving as a complement to the self-attention mechanism. Experimental results indicate that TransCC outperforms existing methods in segmentation performance, boasting an average Dice coefficient of 0.730 and an average Intersection over Union (IoU) of 0.582. These results underscore the effectiveness of TransCC in CCTA image segmentation.
\end{abstract}


\end{frontmatter}

\section{Introduction}\label{sec1} 
The accurate segmentation of Coronary Computed Tomography Angiography (CCTA) \cite{jawaid2017hybrid} is pivotal in the early diagnosis and treatment of Coronary Heart Disease (CHD) \cite{williams2016use, xu2017beat, 10075508}. CCTA images provide clinicians with a detailed view of the heart and its blood vessels, enabling the identification of CHD-related conditions such as atherosclerosis, stenosis, and blockage. The segmentation process partitions the CCTA image into distinct segments, each representing a different anatomical structure. This not only expedites the detection of CHD but also facilitates a more precise, quantitative assessment of the disease. By enhancing the speed and accuracy of CHD diagnosis, CCTA segmentation improves the overall efficiency of CHD treatment, allowing for timely and effective medical intervention.

Existing methods for CCTA segmentation \cite{song2019coronary} predominantly rely on Convolutional Neural Networks (CNNs) \cite{li2018h,myronenko20183d,DBLP:journals/corr/abs-1908-02182,isensee2021nnu,long2015fully,xu2018mutgan}. However, these methods encounter limitations when attempting to accurately segment CCTA. The coronary vascular structure is complex and subject to noise \cite{song2022two, 9386256}, as each slice contains varying numbers of coronary substructures, and the coronary artery structure is often disrupted by cardiac motion noise and non-vascular structures. This complexity necessitates the capture of contextual information from a global perspective to achieve accurate CCTA segmentation. However, CNNs, characterized by their local operations, scan the input image with small convolution kernels and extract local features. This approach results in suboptimal CCTA segmentation performance due to its limited capacity to capture global information. Therefore, there is an urgent need for a method that can effectively capture global information to enhance the accuracy of CCTA segmentation.

\begin{figure}[!t]
	\centerline{\includegraphics[width=1\linewidth]{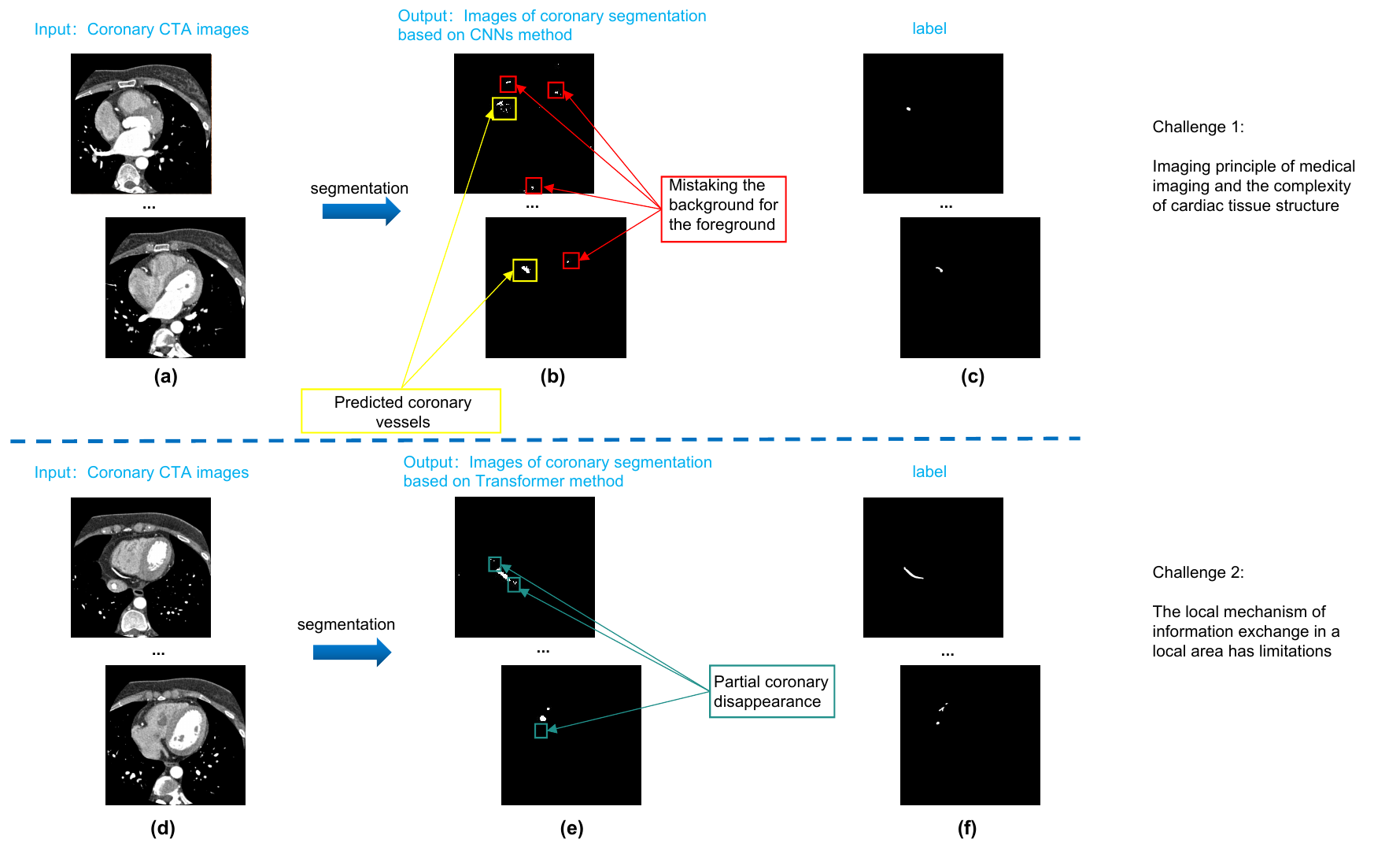}}
	\caption{Example CCTA image. (a), (d) are the original 2D slices. (c), (f) are the Ground Truth of the corresponding slice, and the white area represents the corresponding coronary artery. (b), (e) represent the segmentation results, black represents the background, white represents the predicted blood vessels, and red boxes and red clippings represent wrong predictions.\label{fig1}}
\end{figure}

The Vision Transformer (ViT) \cite{dosovitskiy2020image, wang2018non}  method is currently widely utilized in the field of computer vision. However, due to the small data size inherent in medical imaging, it struggles to address the challenges mentioned above. Contemporary transformer methods \cite{DBLP:journals/corr/abs-2111-14791,liu2021swin,DBLP:journals/corr/abs-2102-04306} employ a self-attention mechanism to model long-range dependencies, achieving superior performance given large-scale datasets. These transformers divide the input image into fixed-size patches and use a self-attention mechanism to model the global context. This approach, however, often results in the segmentation target structure being significantly fragmented. Such division makes it challenging for the network to capture complete features and accurately segment the target, potentially leading to the loss of some latent basic structural features, such as corners and edges. Figure \ref{fig1} illustrates representative CCTA slices. Images (c) and (f) show expert-performed segmentation results, while (b) and (e) are the corresponding initial slices. The complexity of cardiac tissue structure and imaging principles disrupt the coronary vascular structure, making segmentation using visual transformers challenging. For example, Figure 1(b) shows a CNNs-based method misidentifying background pixels due to lack of long-term dependency. Figure 1(e) shows the disruption of the local vascular structure, indicating that fixed-patch segmentation fragments the target structure. Furthermore, the standard transformer relies on a large amount of data to support its linear function fitting after flattening the input image into a vector. However, medical images differ from natural data, and the data sample capacity is relatively small. This makes it challenging to efficiently train a medical transformer.

In this study, we present TransCC, a novel architecture that leverages the transformer for accurate CCTA segmentation. TransCC integrates the strengths of CNNs, such as extracting low-level features and enhancing locality and inductive bias, into the transformer architecture, while preserving the transformer's inherent properties like parallel attention and global context. Given the typically small size of CCTA data, the CNNs significantly reduce the sizes of the transformer's input and linear functions. This results in a substantial improvement in segmentation performance and robustness without increasing the model's complexity. Our method primarily comprises two components: 1) A Feature Extraction Interaction (FIE) module, which addresses the issue of local structure disruption in the transformer image patches. This module eliminates the need for predefined fixed image blocks, enabling adaptive segmentation and feature extraction from the image base. 2) A Multilayer Enhanced Perceptron (MEP) module, which resolves the loss of locality in the spatial dimension when the transformer establishes global correlations between tokens. The feedforward network in each encoder is replaced with a multilayer enhanced perceptron to augment the encoder's ability to capture local context.

In summary, our primary contributions are as follows:
\begin{itemize}
    \item We introduce TransCC, a method designed to effectively amalgamate local and global information for accurate CCTA segmentation. To the best of our knowledge, this represents the first attempt to employ the transformer method for CCTA segmentation.
    \item We propose the Feature Extraction Interaction (FIE) module. This module addresses the challenge of requiring large amounts of pre-training data for effective feature extraction in patches, a common issue in network frameworks based on the visual transformer.
    \item We introduce the Multilayer Enhanced Perceptron (MEP) module. This module resolves the issue of the transformer's lack of locality when applied to CCTA segmentation. It enhances segmentation accuracy and improves the ability to capture local context.
\end{itemize}

The remainder of this article is structured as follows: Section 2 introduces recent related work and methods for CCTA image segmentation. Section 3 provides a detailed description of our segmentation network framework, including the principles and applications of the fundamental modules. Section 4 presents the segmentation results, encompassing comparative experiments and ablation studies. Finally, Sections 5 and 6 conclude the paper and propose future research directions, respectively.

\section{Related work}\label{sec2} 
In this section, we present a comprehensive review of both traditional and recent methods for medical image segmentation along with the approaches that have used transformers to improve it.
Most of the transformer-based methods typically utilize a hybrid architecture of convolution and self-attention\cite{DBLP:conf/nips/VaswaniSPUJGKP17, xu2021video, zhang2023heuristic, xu2020contrast, 9924606}. 
We categorize these approaches into two categories: existing medical image segmentation methods and transformer-based segmentation methods.

Although the abundance of literature on the use of transformer-based methods for medical image segmentation, our focus is primarily on the automatic segmentation of coronary arteries due to its relevance to our work. Our study on cardiovascular image segmentation reveals the absence of a transformer-based method for coronal artery segmentation. This absence may be attributed to the unique position and function of coronary vessels in the cardiovascular structure.

\subsection{Existing methods for segmentation of medical images}
\par Traditional medical image segmentation methods primarily comprise of model-based\cite{holden2007review,zhu2009voles,xu2018direct,xu2019novel} and region-based methods\cite{he2015delving, xu2017direct}. The rapid development of deep convolutional neural networks (CNNs) has led to the widespread adoption of fully convolutional neural networks, specifically encoder-decoder architectures like U-Net\cite{ronneberger2015u}, for many medical image segmentation tasks. These networks have achieved state-of-the-art performance due to their exceptional representational capabilities.
Despite the desirable results produced by these networks, CNNs have significant limitations in modelling global correlations and acquiring global contexts, resulting in serious implications for challenging segmentation tasks.

\subsection{Application of Vision Transformer's Method in Medical Image segmentation}
\par   The Vision Transformer (ViT)\cite{dosovitskiy2020image} has recently demonstrated that ViT can achieve state-of-the-art performance on image classification tasks when the data volume is large enough (i.e., JFT-300M, ImageNet-22k)\cite{sun2017revisiting,xu2022contrast,xu2023spatiotemporal}. In contrast to CNNs-based methods \cite{ronneberger2015u,zhou2018unet++,long2015fully,xu2020segmentation}, ViT is distinct in its utilization of a self-attention mechanism to capture global correlations. This differs from CNNs-based methods that only focus on local features and fail to take into account the global context information, resulting in improved performance. ViT divides input images into patches and considers these patches as words which are the input to the self-attention mechanism. However, the widespread application of ViT is limited by its reliance on large-scale datasets which pose limitations on its use in scenarios with inadequate computing resources or labeled training data. DeiT \cite{touvron2021training} has modified the transformer's layered structure to handle high-resolution images, and it simulates the output of CNNs teachers using knowledge distillation to achieve satisfactory results in ImageNet. However, a potential obstacle is that training high-performance CNNs models are necessary, leading to computational burdens. MedT \cite{valanarasu2021medical} introduced a gated axial attention model, which extends the existing convolutional neural network architecture by adding more control to the self-attention mechanism. Additionally, MedT proposed a local-global (LoGo) training strategy, where it applies a shallower global branch and a deep local branch to extract features from image patches. ReSTR\cite{DBLP:conf/miccai/KarimiVG21} developed a 3D segmentation model based on transformers and removed the need for the convolution operation. The approach is to partition local volumes into 3D blocks, which later get flattened and transformed into 1D sequences, passed to a ViT-like backbone to extract feature representations. Swin-unet \cite{DBLP:journals/corr/abs-2105-05537} constructed a U-shaped transformer segmentation model based on the novel transformer module proposed in \cite{liu2021swin}. This encoder essentially achieves self-attention from local to global, marking an important improvement over \cite{liu2021swin}. DS-TransUNet \cite{lin2022ds} proposes a new segmentation model, which is further extended from the research of Swin-unet by firstly introducing an encoder to process multi-scale input, and secondly introducing a fusion module. Finally the model establishes global dependencies between features at different scales through a self-attention mechanism. Although they have all made improvements to a certain extent, they have not discussed how to effectively combine convolution and self-attention mechanisms to form an optimal segmentation network.

It has been recently shown that utilizing a hybrid architecture that combines Transformer and convolution helps to leverage the strengths of both architectures. Levit-Unet \cite{xu2021levit} explored the equilibrium between the transformer framework and convolutional networks. TransBTS \cite{DBLP:conf/miccai/WangCDYZL21} initially employs 3D convnets to extract volumetric spatial features from the input image, generating the corresponding hierarchical representation. Next, it utilizes the extracted features as input for the transformer to model the global features. Finally, it applies the convolution-based decoder to implement the sampling for the reconstruction of segmentation targets. For the first time, TransUNet \cite{DBLP:journals/corr/abs-2102-04306} merges transformer and convolution to enhance the segmentation performance of medical images. The network framework utilizes the convolutional network as a feature extractor to create a feature map from the input image. Afterward, it applies patch embedding to the bottleneck feature map patch instead of directly applying image embedding to the original image. TransFuse \cite{zhang2021transfuse} introduces a BiFusion module to blend characteristics from two branches of the segmentation network, wherein one is a convolutional network-based encoder, and the other is a transformer-based segmentation network. In contrast to TransUNet \cite{DBLP:journals/corr/abs-2102-04306}, these methods predominantly employ the self-attention mechanism to the input embedding layer to improve the segmentation model of 2D images. Although these methods study how to effectively combine the convolutional neural network and the transformer framework, however, they have a common problem: The convolutional network is used as the main body, and transformer is further applied on this basis to capture long-term dependencies, which may cause the advantages of transformers to not be fully utilized. However, medical images differ from natural images as they have smaller data samples, and current transformer methods \cite{DBLP:journals/corr/abs-2105-05537,han2021transformer,zhang2021transfuse}  are not well suited for training on such a limited dataset. The models demand large-scale data for adequate training purposes \cite{dosovitskiy2020image}. Thus, the existing Transformer-based methods are still under-developed concerning medical image segmentation tasks.

\par Our TransCC method is based on the UNETR \cite{hatamizadeh2022unetr} method, which is currently one of the most efficient and highest-performing techniques utilizing the ViT method. In this architecture, low-level features are extracted using CNNs, followed by modeling of global context through the Transformer. The advantage of this architecture is dedicated to combining CNNs and the Transformer for the interaction of local features and global correlations. Therefore, the fusion of CNNs-based shallow encoding and transformer-based global modeling can successfully address the lack of training data and local modeling.

\section{Methodlogy}
Deep learning networks are intended to model real-world specific issues using mathematical models that permit solving similar problems in the domain. More specifically, representing image data abstractly involves mapping it through a non-linear function f():x → y, where x and y stand for the input and output spaces respectively, and f() denotes the function utilized for such transformation.
An important part of the deep learning network is the convolution layer, and the formula of the convolution layer can be expressed as:
\begin{eqnarray}
  a_{m}^{l+1} =f\left ( \sum_{n=1}^{N} W_{mn}^{l} *a_{n}^{l}+b_{m}^{l+1}  \right ) 
\end{eqnarray}
\par where $N$ represents the number of feature maps in the $l$th layer, $a_{n}^{l}$ is the $n$th feature of the $l$th layer, $a_{m}^{l+1}$ is the $m$th feature map of the $(l+1)$th layer, $W_{mn}^{l}$ is the convolution kernel from the $n$th feature map of the $l$th layer to the $m$th feature map of the $(l+1)$th layer, and $b_{m}^{l+1}$ is the corresponding bias vector\cite{qi2019automated}. In the process of feature mapping, $*$ represents the convolution operation.

\begin{figure}[!t]
	\centering
	\includegraphics[width=1\linewidth]{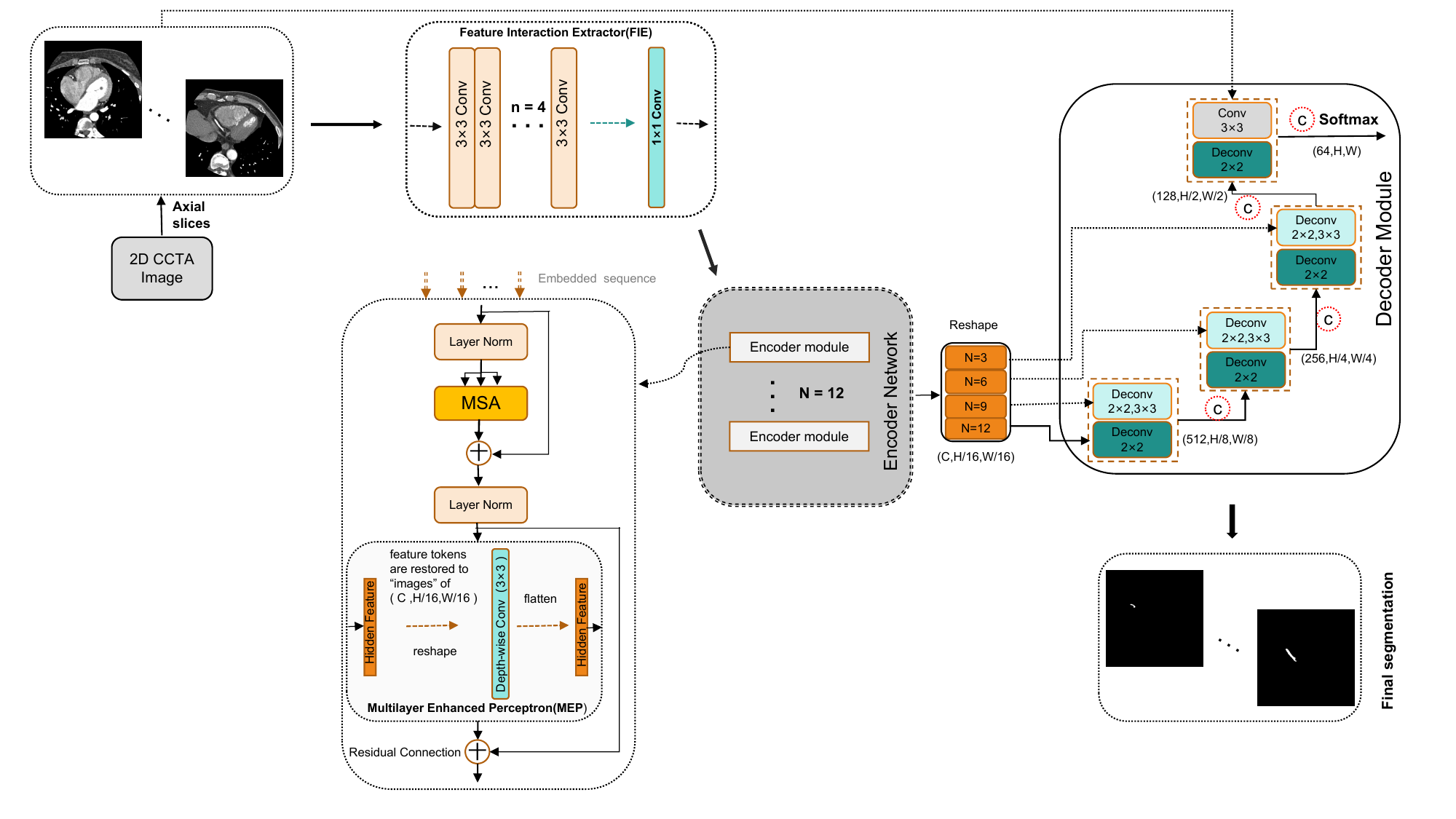}\\
	\caption{ Illustration of TransCC network, which consists of the FIE module, the encoder network, and the decoder module. Among them, the encoder module is mainly composed of a Multi-head Self-Attention and a MEP module, and the encoder modules are stacked in sequence to form an encoder network.\label{fig2}}
\end{figure}

The proposed segmentation network, TransCC, is illustrated in Figure \ref{fig2} and consists of the FIE module, encoder network, and decoder module.
The encoder network is composed of stacked encoder modules, with each module containing a Multi-head Self-Attention and a MEP module.Specifically, TransCC is mainly composed of a Feature Interaction Extraction module (Sect.3.1), a transformer-based encoder module (Sect.3.2), and a decoder module (Sect.3.3). 
The FIE module effectively extracts and retains the semantic information in the input image patch. The transformer-based encoder module employs a self-attention mechanism to model global contextual information and increases locality in spatial dimensions, leveraging our proposed MEP module. Finally, the decoder module uses upsampling and skips connections to obtain high-resolution segmentation results gradually.

\subsection{Semantic Feature Extraction}
\begin{figure}[!t]
	\centering
	\includegraphics[width=0.8\linewidth]{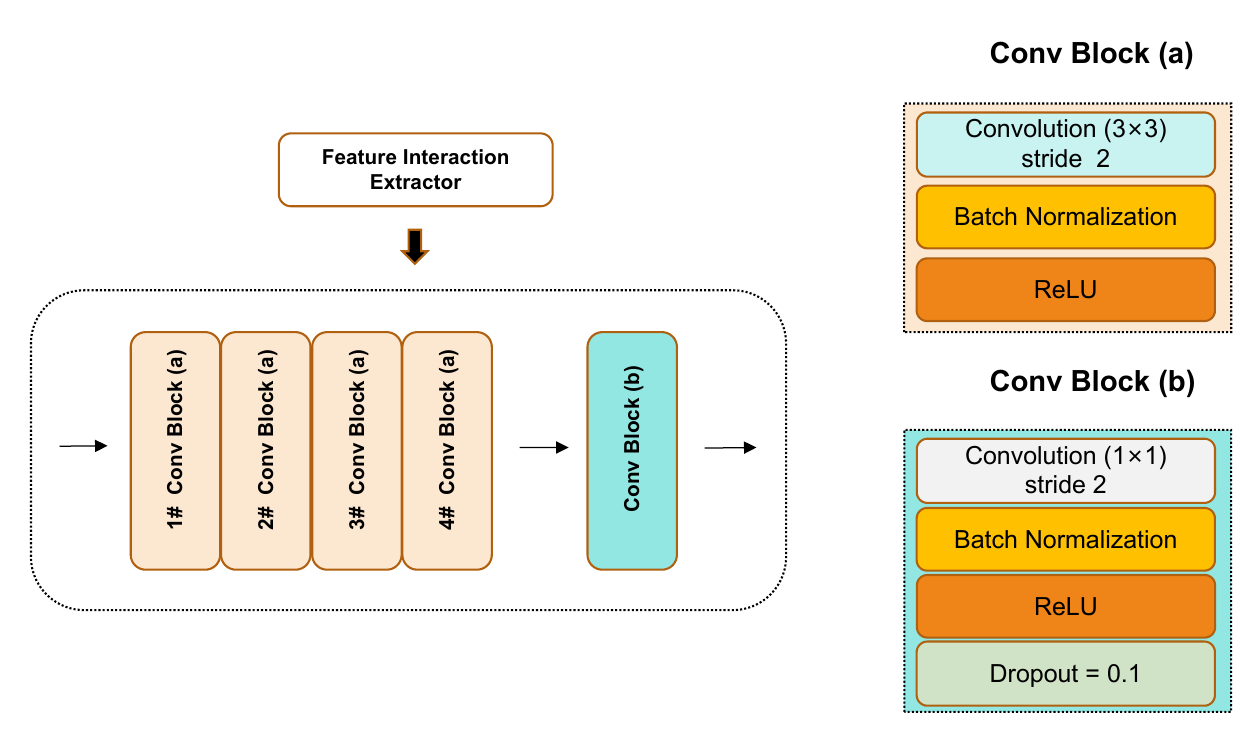}\\
	\caption{ Illustration of FIE module, which mainly consists of 5 stacked convolutional blocks, the input size of our FIE module is set to 224 \(\times\) 224 ( Height \(\times\) Width ) of the initial CCTA image.\label{fig3}}
\end{figure}

Our FIE module combines multi-layer convolutions to effectively explore and preserve semantic information in patches, facilitating the subsequent transformer-based network framework to model local context in both depth and spatial dimensions, thereby improving segmentation effectiveness. We use multiple stacked convolutions to extract more precise spatial information by encoding pixel-level spatial data, which is undoubtedly more accurate than mere patch position encoding.Additionally, stacked convolutions not only prevent semantic overfitting but also enhance the segmentation efficiency of the model. As depicted in Figure \ref{fig3}, the module comprises two parts. The first module has four convolutional layers with a [$3\times3$] kernel size, each followed by a RELU activation function and a layer normalization (i.e., BatchNorm) layer. 
The second module consists of a convolutional layer with a [$1\times1$] kernel size, followed by a BatchNorm layer and a RELU activation function. We also include a Dropout layer at the end to improve the generalization ability of the model. The input resolution of our FIE module is $H\times W$ ($H=224$, $W=224$). The FIE module is a pixel-level encoder network\cite{xu2021synthesis,li2021localvit}, which makes each input image encode target features and obtain more semantically meaningful information. The FIE module not only avoids the destruction of semantic information by traditional patch segmentation methods but also improves the segmentation performance of the model.

The input to our FIE module is the CCTA image $ x \in R^{C\times W \times H}$, where $H \times W$ represent the height and width of the input image respectively, $C$ is the number of channels. Formally, for a given 2D image or feature map token map from the previous layer $ x_{i-1}$, as input to the $i$ th layer, we learn a channel dimension that is the f( ) function of C, mapping the $ x_{i-1}$ of the previous layer to a new f($ x_{i-1}$), where f( ) is a kernel size $K \times K$ (k = 3), and the stride is s (s = 2), padding is a two-dimensional convolution operation of p(p=1) (used to handle boundary conditions). The height and width of the new token map f($ x_{i-1}$) are:
\begin{eqnarray}
	H_{i}=\left \lfloor \frac{H_{i-1}+2p-s}{s-0} +1 \right \rfloor ,W_{i}=\left \lfloor \frac{W_{i-1}+2p-s}{s-0} +1 \right \rfloor 
\end{eqnarray}

The last convolutional module in the FIE module transforms the input map into a higher-level tensor, denoted by $ E_{p} \in R^{C_{f}\times \frac{W}{p}  \times \frac{H}{p} } $. 
To retain the location information in each embedded patch, we incorporate location-specific embedding as follows:
\begin{eqnarray}
	Z_{0} =[X_{f}^{1} E ; X_{f}^{2} E ; \ldots ; X_{f}^{L}E] +E_{p}
\end{eqnarray}
where $E$ represents the linear projection and $E_{p} \in R^{C_{f}\times \frac{W}{p}  \times \frac{H}{p} } $ represents the position embedding. We use $Z_{0}$ as the input to the encoder to further learn the global correlation of the global receptive field.

\subsection{Encoder network}
\subsubsection{ViT-based approach for learning global features}
In the Vision Transformer approach, the Transformer\cite{dosovitskiy2020image,DBLP:journals/corr/abs-2102-04306,DBLP:conf/miccai/WangCDYZL21} encoder consists mainly of a Multi-head Self-Attention (MSA)\cite{vaswani2017attention} module and a Feed-Forward Network (FFN) module (Equations (3) and (4)). The Transformer structure comprises $T$ stacked Transformer-based encoders, where the output of the prior encoder forms the input of the subsequent encoder. The output of layer $i$ $(i\in[1,2,...,L])$ can be calculated as follows:
\begin{eqnarray}
	Z_{i}^{'} = \mathrm{MSA}(Norm(Z_{i-1})) + Z_{i-1}
\end{eqnarray}

\par Here, $Norm$ represents layer normalization, $Z_{i}$ is the image representation output of the $i$-th transformer encoder, and $i$ serves as the intermediate block identifier. 
Each MSA module (i.e., $\mathrm{MSA}$) comprises $n$ parallel self-attentions which are used to learn the similarity between two features of the input sequence ($Z$) and the representation of the query set ($Q$) and key ($K$). 
The self-attention output is calculated as follows:
\begin{eqnarray}
	SA(Z) = Softmax (\frac{qk^{T } }{\sqrt{C_{n} } } )v
\end{eqnarray}

\par where $Z$ represents the value of the query set, and $C_{n}$ represents the scale factor. The Feed-Forward Network module consists of two fully connected layers that perform nonlinear transformation and expansion of the input embedding tokens to improve the tokens' representation ability.
\begin{eqnarray}
	Z_i = FFN(Norm(Z_{i-1})) + Z_{i-1}
\end{eqnarray}

\begin{figure}[!t]
	\centering
	\includegraphics[width=0.8\linewidth]{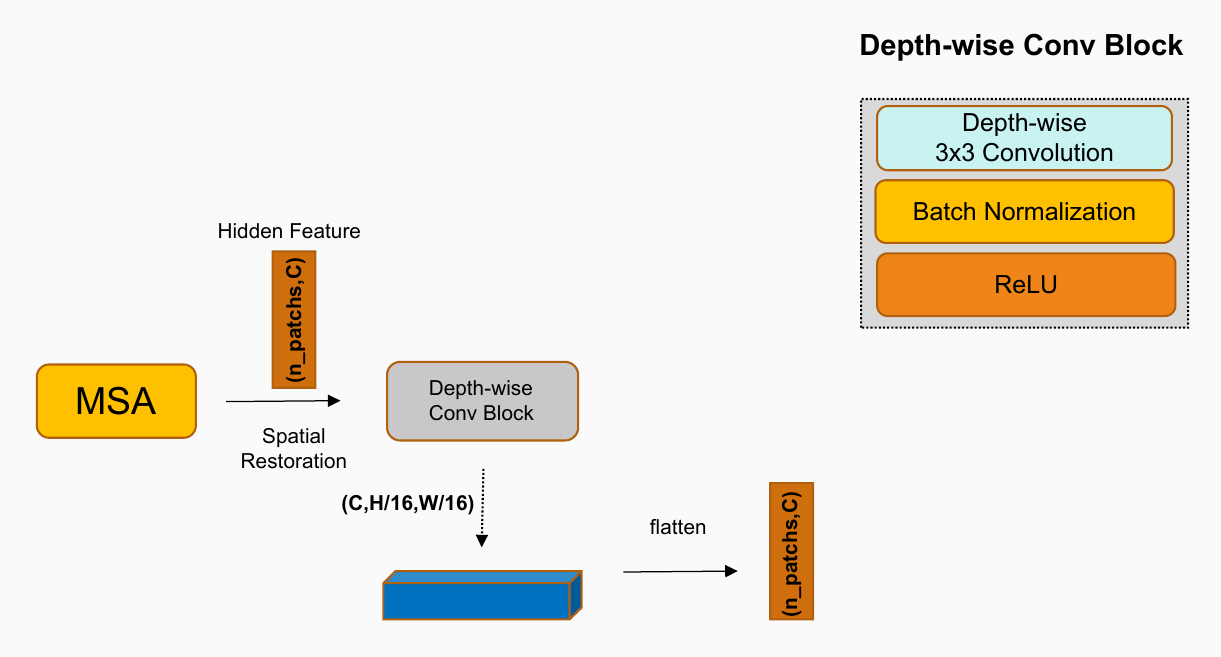}\\
	\caption{Illustration of MEP module. First, the patch tokens received from the MSA module are linearly projected into a higher-dimensional feature space. Second, we restore it to an "image" of $(D,H/16,W/16)$ in the spatial dimension Representation and depth convolution are performed on the reshaped image representation. Finally, the obtained feature representation is flattened and reduced to restore the dimension before the input.\label{fig4}}
\end{figure}

\subsubsection{Combining local transformers structure for Learning global features}
As a supplement to the MSA module, the multi-layer enhanced perceptron module can perform nonlinear transformation and expansion on the input embedding tokens to enhance the tokens' representation ability while addressing the limited ability of the feedforward network in the transformer to leverage the local context. 
As shown in Figure \ref{fig4}, we linearly project the patch labels received by the MSA module into a higher-dimensional feature space, then restore it to an image representation in the spatial dimension, perform depthwise convolution on the reshaped image representation, and finally apply the resulting feature Indicating flattening reduction. Specifically, firstly, for the tokens $Z_{i}^{'}$  (image representations) obtained by the previous MSA module, we use a trainable linear projection to embedding the vectorized patch tokens into higher-dimensional $ Z_{l1}^{'}$, where $\gamma$ is the hidden dimension expansion rate; secondly, we reshape the obtained patch tokens into a 2D feature map, that is, restore the patch tokens to $Z_{p}^{'} \in R^{(C_{f} \times \gamma)\times \frac{W}{p} \times \frac{H}{p}}$ structure, applying a depth-wise convolution with a kernel size of [$3 \times 3$] to the obtained features to obtain local context information has been $ Z_{d}^{'} \in R^{(C_{f} \times  \gamma)\times \frac{W}{p} \times \frac{H}{p}} $. 
The resulting $Z_{d}^{'}$ is then tiled as $Z_{l2}^{'} \in R^{(C_{f}\times\gamma)\times\frac{W\times{H}}{p^{2}}}$ and restored by another linear projection layer the number of dimensions so that it can match the dimensions of the input channel. 
We also employ GELU as the activation function following each linear/convolutional layer. The calculation method is expressed as:
\begin{eqnarray}
	Z_{l1}^{'} = Linear1(Norm(Z_{i}^{'} )),
\end{eqnarray}
\begin{eqnarray}
	Z_{p}^{'} = Rearrange \-Restore(Z_{l1}^{'} ),
\end{eqnarray}
\begin{eqnarray}
	Z_{d}^{'} = GELU(BN(DwConv(Z_{l1}^{'} ))),
\end{eqnarray}
\begin{eqnarray}
	Z_{i2}^{'} = Linear2(Flatten((Z_{i}^{'} )),
\end{eqnarray}
\begin{eqnarray}
	Z_{i} = Z_{i2}^{'} +Z_{i}^{'}
\end{eqnarray}

\subsection{Decoder module}
In the decoder portion of the model, we implemented 2D CNN decoders for feature upsampling to achieve high-resolution feature maps. By combining the features of the encoder and last decoder, we were able to capture both semantic and fine-grained information more effectively, resulting in a richer set of spatial information being obtained.

\par Inspired by the Unet-based network architecture, we fuse features from multiple resolutions of the encoder and the last decoder. We extract the image sequence representation $Z_{i} $ ( i $\in$ [3,6,9,12] ) from the encoder's output, where the size is $ C_{fi} \times \frac{W\times H}{p^{2} } $, and reshape it into a tensor of $ C_{fi} \times \frac{W}{p} \times \frac{H}{p} $, since the image is represented as the output of the encoder(the upper right part of Figure 1) has a feature size of $ C_{f}$(that is, when the image block is embedded to the embedding size before the encoder). The reshaped tensors are projected from the embedding space into the input space using consecutive $3 \times 3$ convolutional blocks before batch normalization.

\par At the encoder bottleneck, which is the output of the last layer of the transformer (i.e., $Z_{12}$ representation of the image), the reshaped feature representation is fed to a deconvolutional layer that upscales the image resolution by a factor of 2. This process allows for the adjusted feature map to be concatenated with the reshaped feature map from the previous stage transformer for fusion (e.g.. $ Z_{12}$). The fused feature maps are then provided as input to consecutive 3$\times$3 convolutional blocks, followed by upscaling with a deconvolution layer. Subsequently, the remaining layers are repeated through this process until the initial image input resolution is reached. Finally, the output is directed to a 1$\times$1 convolutional layer with a softmax activation function, resulting in the final pixel-wise semantic prediction.

\section{Experimental setup} 
This section first presents the dataset information utilized in the experiments. Next, it provides the experimental configuration, implementation details, evaluation metrics, and experimental results. Finally, a series of ablation studies were conducted to systematically analyze the effectiveness and impact of each component in the proposed segmentation network.

\subsection{Datasets details} 
\begin{figure}[!t]
	\centerline{\includegraphics[width=0.45\linewidth]{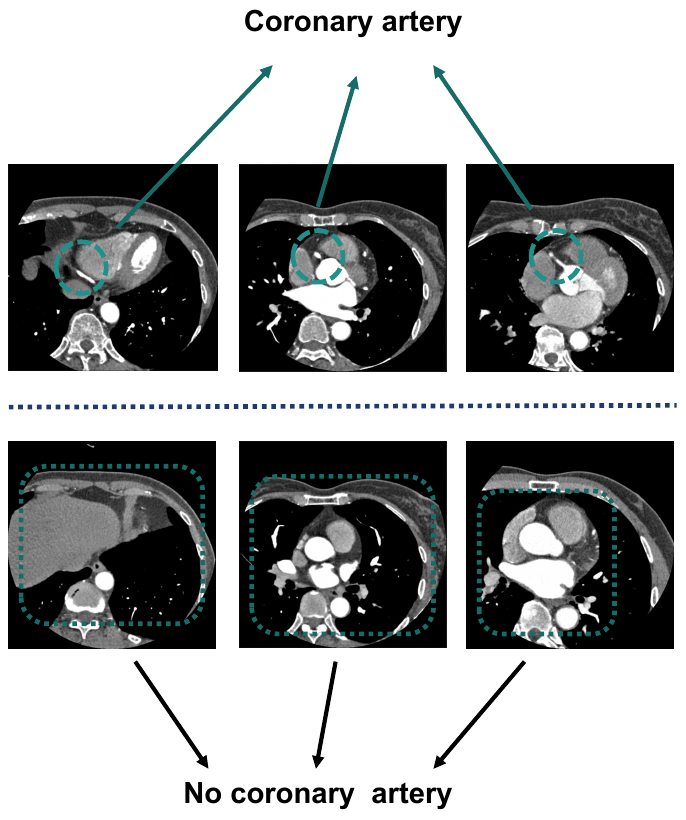}}
	\caption{The original image slice of CCTA is presented. The upper part of the dotted line indicates the presence of the coronary artery, whereas the lower part does not. This study only extracted image data from CCTA.\label{fig5}}
\end{figure}

Our evaluation method utilizes a comprehensive set of CCTA datasets. Each dataset consists of between 250 and 320 slices. Due to the completeness of the CCTA scan images, sequence loss is observed in the initial and final portions of the full slice sequence, as illustrated in Figure \ref{fig5}. We present a selection of representative slices from the original CCTA.Given the complex structure of coronary vessels and the presence of substantial noise interference, only the upper half of the CCTA is discernible, while the lower half is not. Therefore, this study focuses solely on Computed Tomography Angiography images that include coronary arteries. Through manual screening, a total of 848 CCTA images were selected for this study. Each image measures 600 $\times$ 600 pixels.The CCTA data was divided into a training set, randomly selected from six patients' data, and a test set, comprising the remaining data.

\subsection{Implementation}
We implemented TransCC using PyTorch and trained the model on a NVIDIA 2 Tesla P100 with 16GB GPU.
Our network was trained through Binary Cross-Entropy (BCE) loss between the predicted and ground-truth voxels.
Achieving balance between foreground and background voxels is possible without weight assignment to different class samples. Note that we use the BCE loss to be consistent with the baseline to ensure fairness of the experiments. BCE loss is defined as follows:
\begin{eqnarray}
	&l_{n} = - w_{n}[y_{n}\log_{}{x_{n}} +(1-y_{n})\log_{}({1-x_{n}}) ],\\
	&{\Huge \ell } {\tiny bce} ={\Huge \ell }=[{l_{1} ,l_{2},...l_{N}}]
\end{eqnarray}

\par where $x$, $y$ are the predicted values and the labels, respectively, and $N$ is the batch size (4 in our implementation). 
We employ the Adam optimization algorithm with an initial learning rate of 0.001 for 900 iterations.
We use input patches of size 16 \(\times\) 16, stacking the number of transformer modules $T=12$, and embedding size $C=768$ as an encoder for TransCC, where the entire training process does not use any pre-trained transformer models (e.g. ViT on ImageNet).

\par For the purpose of baseline comparisons, we conducted experiments on convolutional and transformer-based segmentation methods. Among the convolutional baselines, we comparedU-net\cite{ronneberger2015u}, Unet++\cite{zhou2018unet++}, V-Net\cite{milletari2016v}, Channel-unet\cite{chen2019channel}, Att-unet\cite{DBLP:journals/corr/abs-1804-03999}, Resunet \cite{zhang2018road}, Resunet++\cite{jha2019resunet++}, and R2unet\cite{DBLP:journals/corr/abs-1802-06955}. For the transformer-based baseline, we compared the UNETR\cite{hatamizadeh2022unetr} method. To propose our method, we conducted three steps. Firstly, we introduced the proposed FIE network module into the encoder-based framework of Vision Transformer. Secondly, we introduced our proposed MEP network module, following the same procedure as the previous step. Finally, we replaced the encoder module of the vision transformer with the FIE module and the MEP module as a whole. For the transformer-based method baseline, we used the encoder part of the Vision Transformer and the decoder consistent with our method.

\subsection{Workflow section}
\begin{figure}[!t]
	\centering
	\includegraphics[width=1\linewidth]{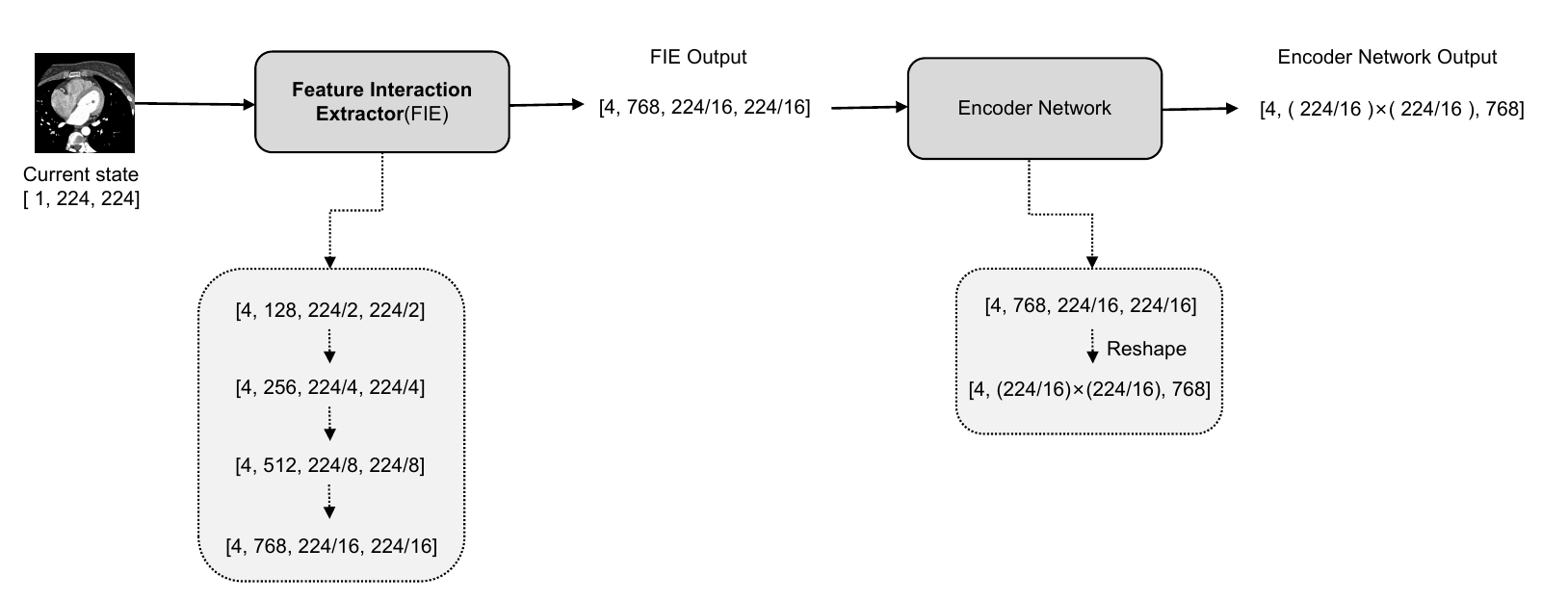}\\
	\caption{ Our method mainly consists of a FIE module and an encoder module to obtain accurate (i.e., pixel-level) spatial information and low-level high-resolution features for an input image. During feature learning, the dimensions of the input are the same as the dimensions of the output.\label{fig6}}
\end{figure}

According to Figure \ref{fig6}, our proposed method mainly consists of a FIE module and an encoder module to obtain accurate (i.e., pixel-level) spatial information and low-level high-resolution features for a input image. 
During feature learning, both the dimensions of the input and the output are the same. 
1) FIE module. 
Input: The current state of the input CCTA image is represented by a tensor of size [4, 224, 224], where 4 refers to the number of channels, and 224 represents the height and width of each frame.  
Output: The FIE module produces an output tensor of size [4, 768, 224/16, 224/16], where 4 indicates that 4 frames are processed simultaneously, 768 denotes the encoding dimension, and 224/16 means the size after encoding by the FIE module.
2) Encoder model. 
Input: As the proposed approach entangles long-term dependencies with local characteristics, the output dimension of the FIE module matches the input dimension of the encoder module. Moreover, the transformer will reshape the input dimension to [4, (224/16) x (224/16), 768], facilitating the similarity calculation of the self-attention mechanism.  
Output: The output of the Encoder network is [4, (224/16) x (224/16), 768], representing the high-level features and the same size of the input tensor. 

\par The two types of output are alternative and collaborative. On the one hand, the embedding of lightweight convolutional layers preserves more accurate spatial information, while, alternatively, connecting transformers to learn context information helps improve the generalization and robustness of learned representations.

\subsection{Evaluation metrics}
\par We utilized five evaluation metrics based on area stacking and surface distance to assess the segmentation performance of our model. The evaluation metrics employed include Dice similarity coefficient (Dice), Intersection over Union (IoU), F1 score, Hausdorff Distance (HD), and Average Symmetric surface Distance (ASD). These metrics were used to independently compute the segmentation results of coronary artery. The formulas for the evaluation metrics are as follows:
\begin{eqnarray}
	Dice(p_{i},g_{i}  )= 2\times \frac{ {\textstyle \sum_{i}^{N}p_{i}g_{i}  } }{ {\textstyle \sum_{i}^{N}p_{i}^{2}+   {\textstyle \sum_{i}^{N}g_{i}^{2} }  } } 
\end{eqnarray}
\begin{eqnarray}
    IoU(p_{i} ,g_{i} ) = \frac{p_{i}\bigcap g_{i}  }{p_{i}\bigcup  g_{i}}  
\end{eqnarray}
\begin{eqnarray}
	F1 = \frac{2T_{P} }{2T_{P}+F_{P}+F_{N}}  
\end{eqnarray}
\begin{eqnarray}
	HD= (max_{i\in P}( max_{i\in G} (d(i,j))),max_{i\in G} (max_{i\in P} (d(i,j)))) 
\end{eqnarray}
\begin{gather}
	B_{seg} =\left \{  \forall p_{1}\in p_{seg}, \text { $closet\_distance$ }\left(p_{1}, p_{2}\right)\mid \exists p_{2}\in g_{gt} \right \}\notag \\
	ASD=\operatorname{mean}\left(\left\{B_{\text {$seg$ }}, B_{gt}\right\}\right)
\end{gather}

\par where $ p_{i} \in P$ represents the binary segmentation result predicted by the network, and  $g_{i}\in G$ represents the ground truth of binary segmentation. 
Dice and IoU are values between 0 and 1, and the goal of the segmentation network is to maximize its value, which is equivalent to increasing the segmentation accuracy. 
$T_{P} $(True positive) represents a positive sample pixel that is correctly predicted,  $F_{P} $ (False positive) represents a wrongly predicted sample pixel as positive, and similarly,  $T_{N} $ (True Negative) and  $T_{N} $ (False Negative)  represent that the prediction is correct and negative, respectively. 
Sample pixels and predictions are wrongly predicted negative samples. 
$d(i,j)$ represents the Euclidean distance, where $(i,j)$ represents the predicted or boundary points on the ground truth segmentation mask boundary. 
$ p_{seg}$ represents the pixels of the boundary of P in the prediction, and $ g_{gt}$ represents the pixels of the boundary in the ground truth.

\begin{table*}[!b]
	\renewcommand\arraystretch{1.2}
	\caption{Quantitative comparison of our method with Convolution-based baselines and visual Transformer-based baselines on CCTA datasets using Average (Dice score, Intersection over Union, F1, Hausdorff Distance, and Average Symmetric surface Distance) five scoring methods evaluate.\label{label1}}
		\resizebox{\textwidth}{!}{
			\begin{tabular}{ccccccc}
				\hline
				\multirow{2}{*}{Methods}&\multirow{2}{*}{Framework}&\multicolumn{5}{c}{CCTA}\\
				\cline{3-7}
				& & Dice ↑& IoU ↑  & F1\_score ↑ & HD ↓ & ASD ↓   \\
				\hline
				\multirow{8}{*}{Convolutional Baselines}& V-Net \cite{milletari2016v}  & 0.647 &0.489 & 0.997  & 8.7311& 7.845 \\
				& U-Net   \cite{ronneberger2015u}      & 0.704 & 0.553& 0.997  & 12.810  & 11.096 \\
				& Att-unet \cite{DBLP:journals/corr/abs-1804-03999}     & 0.685 & 0.531 & 0.998  & 9.746   & 8.800    \\
				& Resunet     \cite{zhang2018road}   & 0.714  & 0.565  &  0.998 & 3.959    & 3.943  \\
				& Resunet++   \cite{jha2019resunet++}  & 0.690 & 0.535  & 0.997 & 12.017 & \textbf{0.065}   \\
				& Unet++  \cite{zhou2018unet++}      & 0.710  & 0.557& 0.998 & 6.458  &  6.012     \\
				& R2unet    \cite{DBLP:journals/corr/abs-1802-06955}     & 0.708& 0.558  & 0.998 & \textbf{3.159 }& 2.696 \\
				& Channel-unet \cite{chen2019channel}  & 0.703 & 0.552 & 0.997 & 6.840& 7.525     \\
				\hline
				Transformer Baesline & UNETR   \cite{hatamizadeh2022unetr}    & 0.655  & 0.497 & 0.979 & 15.596 &  13.047\\
				\hline
				\multirow{3}{*}{Ours}  & FIE    &0.711  &0.560 & 0.998  &  6.458  &  6.012 \\                
				& MEP    &0.698  & 0.546  & 0.998  &5.950  &     4.949    \\
				& \textbf{TransCC}    & \textbf{0.730} &\textbf{0.582} & 0.998     & 6.457 &  6.149  \\
				\hline 
			\end{tabular}
		}
	\end{table*}

\begin{figure}[!t]
	\centering
	\includegraphics[width=1\textwidth]{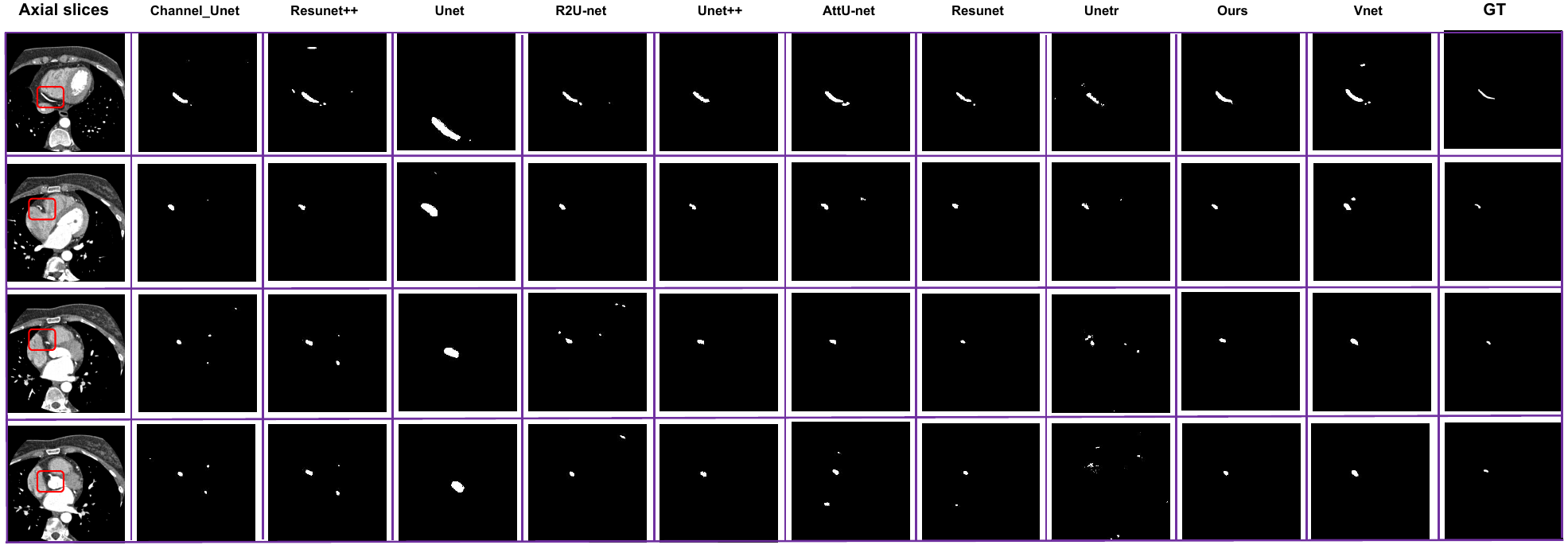}\\
	\caption{ Qualitative results on test image samples from the CCTA datasets. From left to right: input image, segmentation results of different segmentation methods, and Ground Truth. In contrast, our method predicts fewer false positives and preserves finer vessel information. The blue part represents the image background, and the yellow part represents the segmented predicted coronary artery.\label{fig7}}
\end{figure}

\subsection{Experimental results}
In this section, we present a quantitative and qualitative evaluation of our segmentation network’s performance, comparing our results to those generated by other advanced segmentation methods. Our comprehensive experiments demonstrate that TransCC produces high-quality segmented CCTA images, achieving a Dice coefficient and Intersection over Union (IoU) of 0.730 and 0.582, respectively, for automatic coronary artery segmentation. These results underscore the potential of TransCC as a transformer-based method for medical image segmentation.

\subsubsection{Comparison with several related works }
For quantitative analysis, we primarily use the Dice coefficient and IoU score to evaluate the models. Initially, we implemented the classic segmentation networks U-Net\cite{ronneberger2015u} and Resunet\cite{zhang2018road}, and an efficient segmentation network based on transformer along with six other efficient segmentation methods. The quantitative results are shown in Table \ref{label1}. The analysis revealed that among all the compared methods, the transformer-based segmentation method performed inferior relative to the convolution-based baseline. The inferior performance is attributable to the inadequacy of the transformer-based segmentation method in the case of a small CCTA dataset. Addressing the inadequacy, we proposed the FIE network module, which improved the performance such that it was superior to some classical segmentation networks, including U-Net\cite{ronneberger2015u} and V-Net\cite{milletari2016v}. Our final TransCC network framework outperforms both FIE and MEP network modules as well as all previous segmentation methods. Our proposed method shows significant improvement over the transformer-based and the best convolutional baselines in the segmentation results of the CCTA dataset. Specifically, our method outperforms the transformer-based baseline by 7.5\% and 8.5\%, and the best convolutional baseline by 1.6\% and 1.7\% in terms of Dice coefficients and IoU scores. This improvement is due to the segmentation algorithm used in our method.

\par We visualize the proposed method and the predicted segmentation results of all the baselines for qualitative analysis. As can be observed from the quantitative results in Figure \ref{fig7}, we sequentially show the 2D image of the input coronary CT image, the segmentation results of different benchmark methods, and the ground truth. Among them, the black and white parts represent the background of the image and the segmentation results of the predicted coronary arteries, respectively. Our proposed method preserves finer vessel information, as we have observed. We illustrate the four-frame angiographic images, along with their corresponding ground truth and the visualized prediction. These angiograms reveal that coronary vascularity is highly complex and variable, particularly in cases where coronary arteries connect to the aorta. Other methods, in rows 2 and 4 of the visualized prediction results, incorrectly predict background pixels as target pixels because the lesions are in close proximity to these background pixels. Our method integrates the locality and long-range dependencies properties efficiently, making our predictions more precise. Therefore, our proposed method is substantially closer to the ground truth than other state-of-the-art methods.

\subsubsection{Ablation study}
The proposed network segmentation is achieved by integrating multiple modules on the encoder-decoder framework. 
Therefore, it is necessary to conduct ablation studies to analyze the effectiveness of various modules on the segmentation performance.To extract richer underlying features and improve the local pixel integrity of the target object, we design an FIE module in the encoder.Self-attention-based Encoders model only the long-range dependencies of tokens. We propose an MEP module to improve the ability to capture local context while modeling global characteristics.For these structural improvements, we proved their effectiveness through the following ablation experiment. We plan to conduct a detailed study on these network module improvements, which focuses on the impact of input image resolution, diverse hidden dimension expansion rates $\gamma$ in the feedforward network, and the performance of each proposed network module on the segmentation.

\begin{figure}[!t]
	\centering
	\includegraphics[width=0.55\linewidth]{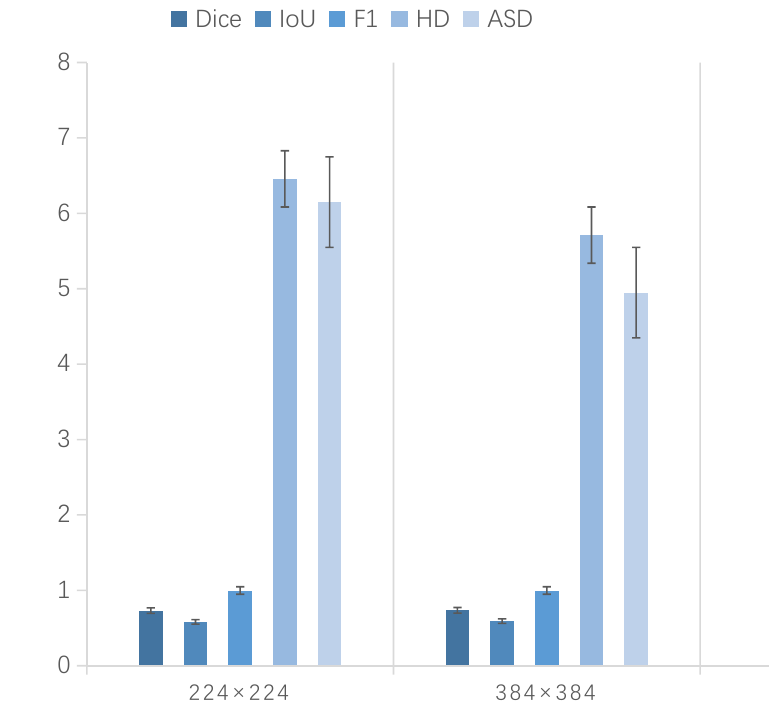}\\
	\caption{Ablation study on the effect of input resolution.\label{fig8}}
\end{figure}

\begin{figure}[!t]
	\centering
	\subfigure{
     \includegraphics[width=0.45\linewidth]{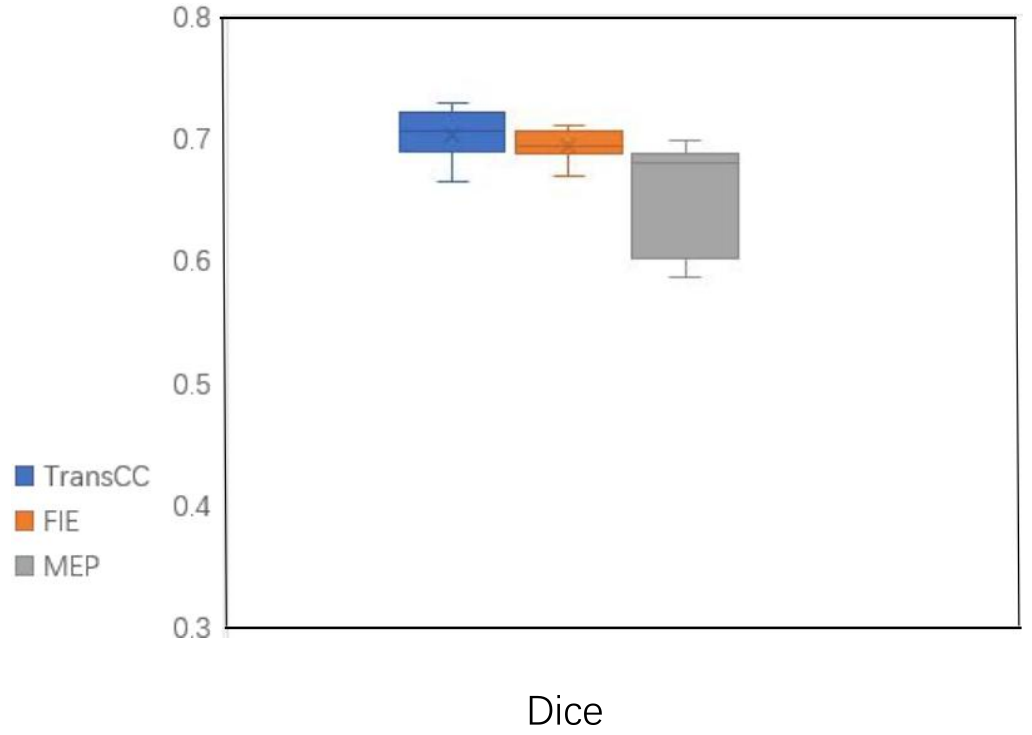}
    }\label{chutian3}
    \quad
    \subfigure{
     \includegraphics[width=0.45\linewidth]{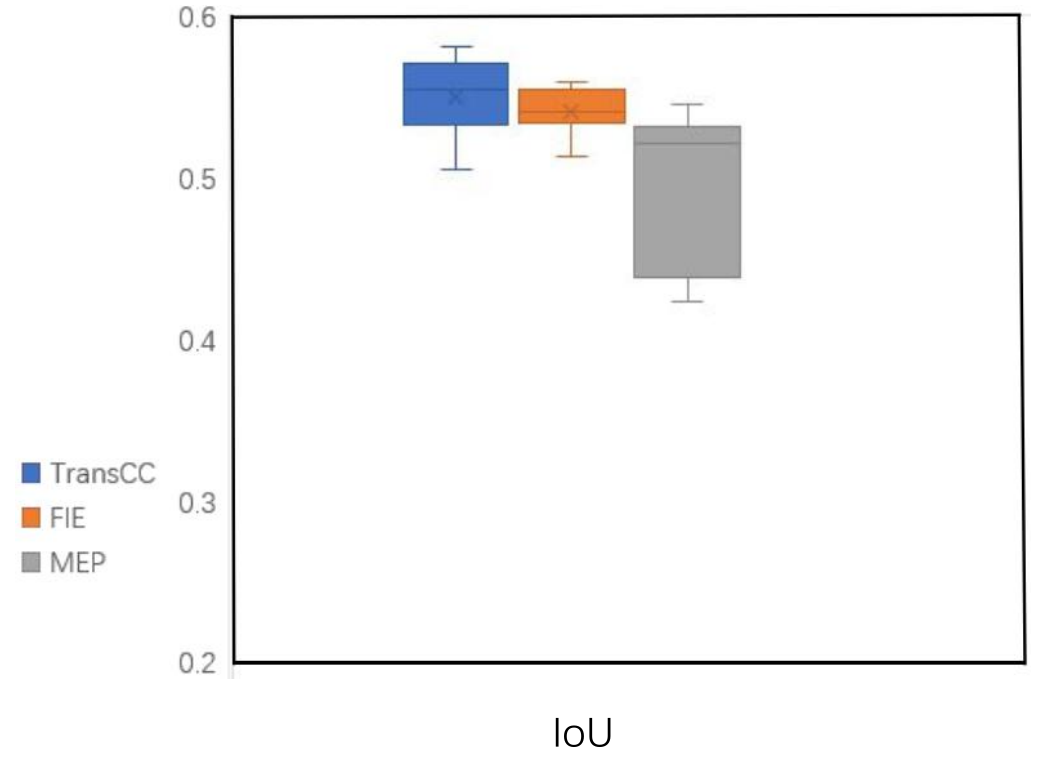}
    }\label{chutian4}
    \quad
	\caption{Ablation study of the Segmentation Network Performance with Different Encoding Modules. \label{fig9}}
	\label{da_chutian}
\end{figure}

\paragraph{Effect of input image resolution}  
During the training phase, TranssCC adjusts its input resolution to 224 $\times$ 224 as default. However, it has been suggested in \cite{dosovitskiy2020image} that an increase in the effective sequence length can robustly enhance segmentation performance. In order to analyze the impact of input resolution on model performance, a higher resolution of 384 $\times$ 384 was also provided as input. Figure \ref{fig8} demonstrates the findings that the average Dice and average IoU improved by 1.1\% and 1.5\%, respectively, when the input resolution was changed from 224 $\times$ 224 to 384 $\times$ 384. However, using 384 $\times$ 384 as input resulted in a smaller effective sequence length than the initial set value, and a further sacrifice in computational cost. In view of the computational cost - including the required FLOPS and parameters of our TransCC, which are 266.79G and 100.43M respectively - we chose to set our input resolution to 224 $\times$ 224 in our experiments.\\

\paragraph{Effect of the proposed network module on the segmentation network} 
Our TransCC segmentation method uses the Transformer as its backbone. We conducted ablation experiments on the network modules to assess their contributions to the segmentation performance of the network as a whole. Figure \ref{fig9} illustrates that the integration of the FIE and MEP modules into the segmentation system alone exceeds some other methods in the CCTA dataset. For instance, when only using the FIE module, the average Dice and average IoU in the segmentation results are 0.711 and 0.560, respectively, which is higher than most convolution-based baselines (such as the U-Net method with average Dice and average IoU of 0.704 and 0.553, respectively). Our TransCC, which is an efficient combination of both modules, achieved the best performance in both average Dice and average IoU. This demonstrates that the proposed FIE and MEP modules are useful in helping the segmentation network learn target pixel features more effectively, thereby improving the performance of the segmentation.\\

\begin{figure}[!t]
	\centering
	\includegraphics[width=0.75\linewidth]{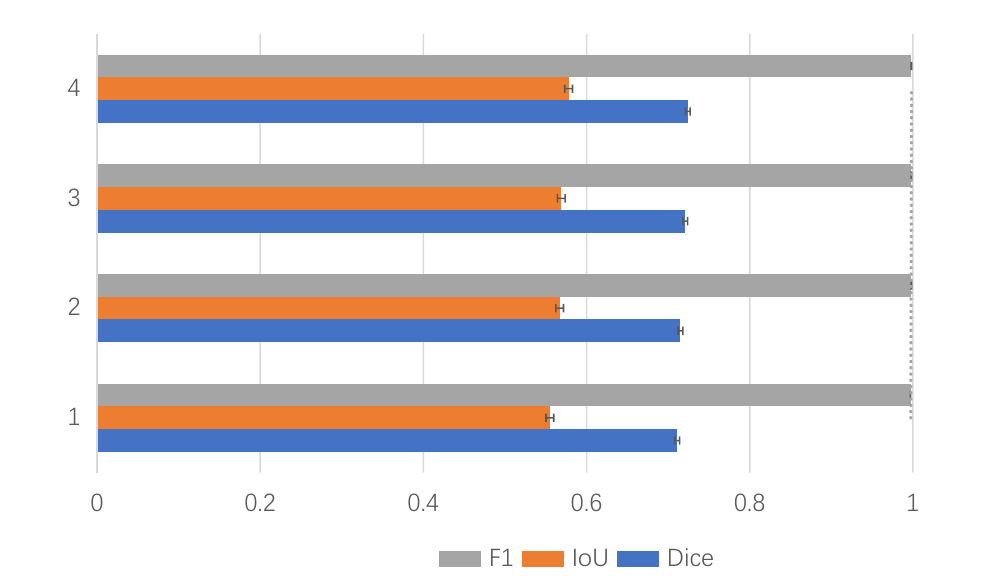}\\
	\caption{Ablation Study of Augmented Scale in Hidden Layers of FIE Networks.\label{fig10}}
\end{figure}

\paragraph{Effect of different hidden dimension expansion rates}  
The encoder of the TransCC model is augmented with a multi-layer perceptron module to extend the feature extraction ability of the feedforward network. Additionally, we analyze the influence of various scaling ratios on the network's segmentation performance. 
As presented in Figure \ref{fig10},  it can be seen that with the increase in the hidden dimension $\gamma$, the segmentation performance of the model improves marginally, denoted by an increase in the average Dice coefficient from 0.715 to 0.730.
This result signifies that changing the hidden dimension ratio $\gamma$ in the MEP module can enhance the performance of our segmentation model.
Thus, we set the hidden dimension ratio $\gamma$ to 4 by default in our experiments.

\section{Limitations and further directions}
The TransCC, as a pioneering work, still has some limitations. 1) Our method only utilizes a single CCTA image as input. Even though we obtained convincing results and demonstrated the potential and feasibility of the method, to improve its performance, it is essential to input a broader range of modalities. In the future, we plan to use multimodal images to enhance the effectiveness of our method. 
2) Our method is relatively computationally expensive due to its transformer architecture. Thus, it inevitably consumes a lot of memory and time. Although the patch embedding training cost reduced with the help of convolutional neural networks, the computational cost of convolution and fully connected matrix calculations for high-resolution images still requires the method to be trained on NVIDIA 2 Tesla P100 16G GPU for nearly seven days. Consequently, in our future work, we aim to employ advanced compression techniques to reduce costs and improve efficiency while maintaining performance.

\section{Conclusion}
This work explores a medical image segmentation method using a system architecture based on transformer encoders. We propose a new segmentation framework which involves incorporation of 2D Convolutional Neural Networks (CNNs) into transformer architecture specifically for Coronary Computed Tomography Angiography (CCTA) segmentation. The ultimate architecture, TransCC, inherits the ability of transformers to model global contextual information in order to aid in learning high-level features; in addition, it utilizes CNNs to learn local semantic correlations to obtain more refined low-level features. Experiments conducted on CCTA datasets successfully demonstrate the effectiveness of TransCC.

\section*{Acknowledgments}
This work was supported in part by The National Natural Science Foundation of China (62106001, U1908211), The University Synergy Innovation Program of Anhui Province under Grant (GXXT-2021-007), and The Anhui Provincial Natural Science Foundation (2208085Y19).




\bibliographystyle{elsarticle-harv} 
\biboptions{sort&compress}
\bibliography{elsarticle-template-harv.bib}

\end{document}